# Scalable quality control on processing of large diffusion-weighted and structural magnetic resonance imaging datasets

Michael E. Kim[a*], Chenyu Gao[b], Karthik Ramadass[a,b], Praitayini Kanakaraj[a], Nancy R. Newlin[a], Gaurav Rudravaram[b], Kurt G. Schilling[c], Blake E. Dewey[d], David A. Bennett[e], Sid O'Bryant[f,g], Robert C. Barber[g], Derek Archer[h,i], Timothy J. Hohman[h,i], Shunxing Bao[b], Zhiyuan Li[b], Bennett A. Landman[a,b,c,j,k], Nazirah Mohd Khairi[b], The Alzheimer's Disease Neuroimaging Initiative[A], The HABS-HD Study Team[B]

## ABSTRACT

Proper quality control (QC) is time consuming when working with large-scale medical imaging datasets, yet necessary, as poor-quality data can lead to erroneous conclusions or poorly trained machine learning models. Most efforts to reduce data QC time rely on outlier detection, which cannot capture every instance of algorithm failure. Thus, there is a need to visually inspect every output of data processing pipelines in a scalable manner. We design a QC pipeline that allows for low time cost and effort across a team setting for a large database of diffusion-weighted and structural magnetic resonance images. Our proposed method satisfies the following design criteria: 1.) a consistent way to perform and manage quality control across a team of researchers, 2.) quick visualization of preprocessed data that minimizes the effort and time spent on the QC process without compromising the condition/caliber of the QC, and 3.) a way to aggregate QC results across pipelines and datasets that can be easily shared. In addition to meeting these design criteria, we also provide information on what a "successful" output should be and common occurrences of algorithm failures for various processing pipelines. Our method reduces the time spent on QC by a factor of over 20 when compared to naïvely opening outputs in an image viewer and demonstrate how it can facilitate aggregation and sharing of QC results within a team. While researchers must spend time on robust visual QC of data, there are mechanisms by which the process can be streamlined and efficient.

[a]Vanderbilt University, Department of Computer Science, Nashville, TN, USA

[b]Vanderbilt University, Department of Electrical and Computer Engineering, Nashville, TN, USA

[c]Vanderbilt University Medical Center, Department of Radiology and Radiological Sciences, Nashville, TN, USA

[d]Department of Neurology, Johns Hopkins University School of Medicine, Baltimore, Maryland, USA

[e]Rush Alzheimer's Disease Center, Rush University Medical Center, Chicago, IL, USA

[f]Institute for Translational Research, University of North Texas Health Science Center, Fort Worth, Texas, USA

[g]Department of Family Medicine, University of North Texas Health Science Center, Fort Worth, TX, USA

[h]Vanderbilt University Medical Center, Vanderbilt Memory and Alzheimer's Center, Nashville, TN, USA

[i]Vanderbilt University Medical Center, Vanderbilt Genetics Institute, Nashville, TN, USA

[j]Vanderbilt University, Department of Biomedical Engineering, Nashville, TN, USA

[k]Vanderbilt University Institute of Imaging Science, Nashville, TN, USA

investigators can be found at: http://adni.loni.usc.edu/wp-content/uploads/how_to_apply/ADNI_Acknowledgement_List.pdf

[B]HABS-HD MPIs: Sid E O'Bryant, Kristine Yaffe, Arthur Toga, Robert Rissman, & Leigh Johnson; and the HABS-HD Investigators: Meredith Braskie, Kevin King, James R Hall, Melissa Petersen, Raymond Palmer, Robert Barber, Yonggang Shi, Fan Zhang, Rajesh Nandy, Roderick McColl, David Mason, Bradley Christian, Nicole Phillips, Stephanie Large, Joe Lee, Badri Vardarajan, Monica Rivera Mindt, Amrita Cheema, Lisa Barnes, Mark Mapstone, Annie Cohen, Amy Kind, Ozioma Okonkwo, Raul Vintimilla, Zhengyang Zhou, Michael Donohue, Rema Raman, Matthew Borzage, Michelle Mielke, Beau Ances, Ganesh Babulal, Jorge Llibre-Guerra, Carl Hill and Rocky Vig.

*Corresponding Author: michael.kim@vanderbilt.edu

## 1. INTRODUCTION

The rate of medical imaging data availability for research is ever increasing as we continue to integrate technology into the healthcare system (Abhisheka et al., 2024; European Society of Radiology, 2010; Nabrawi & Alanazi, 2023; Siegel, 2021). This influx of imaging data, especially magnetic resonance imaging (MRI) in the field of neuroimaging (LaMontagne et al., 2019; Littlejohns et al., 2020; Lorenzini et al., 2022; Snoek et al., 2021; Weber et al., 2021), allows us to perform large-scale data analyses important for drawing generalizable and reproducible conclusions about human health (Aerts, 2018; Button et al., 2013; S. Liu et al., 2023; Marek et al., 2022) or more robustly training deep learning algorithms (Q. Liu et al., 2020). However, researchers often do not analyze these raw image data directly; rather, image processing pipelines are used to extract derived quantitative metrics for subsequent scientific investigations (Vadmal et al., 2020). The scope of derived data can range from first-order features such as total brain volume (Y. Liu et al., 2022) to complex 3D representations of white matter brain pathways (Wasserthal et al., 2018). Unfortunately, not all these available derived data are suitable for use in scientific research, as the algorithmic outputs may not exactly match the features these pipelines are attempting to capture for any given input scan (Antun et al., 2020). Implementing robust quality control (QC) protocols can help ensure the quality and consistency of data used for research, facilitating more accurate and reproducible scientific breakthroughs (Ducharme et al., 2016) and better machine learning models (Santos, 2023).

There are several considerations for ensuring effective QC of large-scale datasets. First, it quickly becomes infeasible for a single researcher to perform all QC tasks as more data are introduced; thus, QC can become more achievable when it is partitioned to members of the team that process the data. However, there are additional considerations for a team-driven effort, such as the manner in which QC for a pipeline is performed. There are a multitude of freely downloadable image viewers available for researchers, and while most provide the same general functionality, each is usually specialized for a specific data type. For example, FSLeyes (McCarthy, 2024) is specialized for viewing timeseries data and outputs from FSL (Jenkinson et al., 2012) processing, whereas MI-brain (Rheault et al., 2016) is built for visualizing tractography, but both are capable of viewing 4D imaging data in similar ways with differences in presentation to the user and the graphical user interface (GUI). Yet discrepancies in any aspect of the QC process can greatly increase the amount of time spent aggregating results across a team or can result in data being assessed with different standards. Thus, the remaining time available to perform analyses on the data can be greatly impacted and can even result in data of different quality standards being introduced in the dataset, which can impact results and conclusions from scientific studies (Alexander-Bloch et al., 2016; Craig Reynolds et al., 2023; Ducharme et al., 2016; Power et al., 2012).

Further challenges related to the aspect of team QC are how the QC status of data is being reported and the standards by which researchers assess poor or good quality. While a more complex QC method that assesses multiple dimensions of quality can provide more insight into the data (Kindling & Strecker, 2022; Taleb et al., 2021a), such an assessment would greatly increase the amount of time spent performing QC. For example, a metric of quality such as "accuracy" as defined by Kindling et al., Goasdoue et al., and Taleb et al. (Goasdoué et al., 2007; Kindling & Strecker, 2022; Taleb et al., 2021b), could be used in a deeper dive for what makes certain qualities of data more susceptible to algorithm failure.

However, the foremost goal of pipeline QC when handling large-scale MRI data should not be complex judgement of data quality, especially for image processing researchers whose area of expertise does not pertain to MRI scanner systems and software that collect data. No image processing algorithm is perfection incarnate, especially not deep learning-based processing pipelines (Antun et al., 2020). Thus, in QC of this scale, the most important aspect is to assess if the algorithm performed as expected or if it failed to produce a believable output.

To that end, another critical concern is the amount of time spent actually performing QC on the data. Though QC is essential for any research study, it is beneficial to optimize the amount of time spent on the process. As mentioned above, there is a panoply of visualization tools available for looking at data. However, opening individual files in these image viewers is far too time-consuming. Time minimization of QC is a well-researched area in imaging informatics. MRIQC and other works have focused on automated or semi-automated outlier detection using image-derived metrics (Alfaro-Almagro et al., 2018; Esteban et al., 2017; Lorenzini et al., 2022). There have also been efforts to automate QC of outputs from processing pipelines such as segmentation algorithms (Q. Liu et al., 2023; Sunoqrot et al., 2020). While such methods and tools can considerably cut down the time spent on QC, they are techniques for batch QC of data reliant on quantitative summary metrics rather than QC of individual images and outputs. These batch QC methods cannot consistently capture when an algorithm breaks or fails because they rely on summary metrics of the data that could be unrepresentative of the true data quality: one cannot correctly represent quality of a high-dimensional signal such as a 3D MRI using a single, derived quantitative value. Additionally, an algorithm can work as expected, but still flag data as an outlier due to the nature of the original data, even when the quality is good. Thus, proper QC requires visualization of every image or data output to ensure that the algorithm performed as expected while remaining efficient with regard to the amount of time spent. Benhajali et. al proposed a method for quick QC of image registration that involved visualizing each result but was limited only to image registration pipelines (Benhajali et al., 2020).

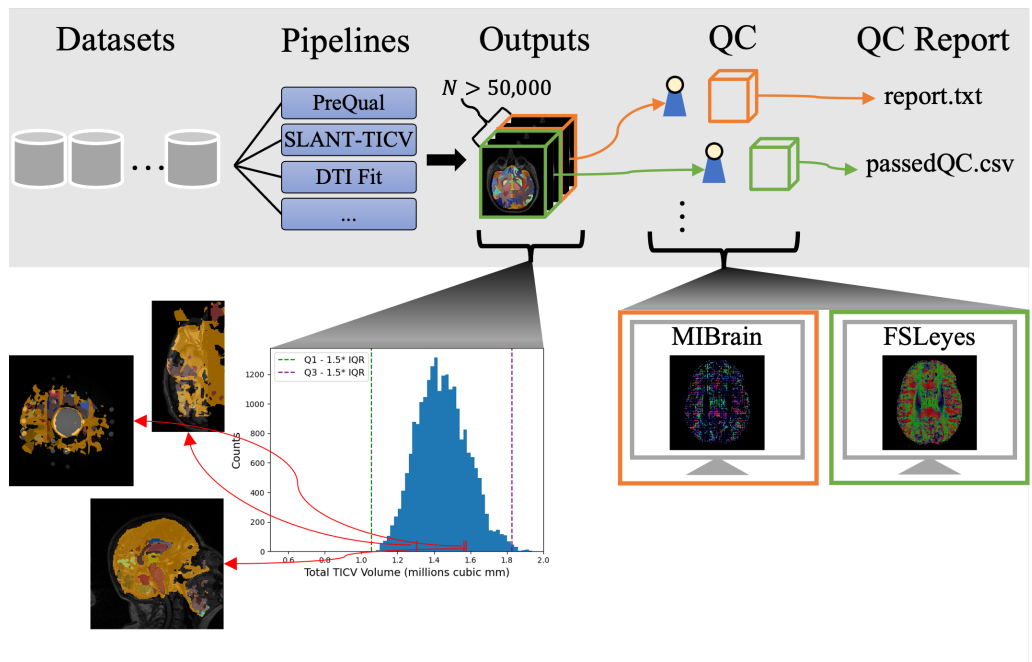

**Fig. 1** When maintaining large neuroimaging datasets with multiple processing pipelines, shallow quality control processes that rely on derived metrics can fail to catch instances of algorithmic failures. However, deep QC processes quickly become unscalable and inefficient as the amount of data available increases due to the required time for mass visualization of outputs. For example, opening 50,000 T1w images separately in an image viewer for deep QC can take over 60 hours if it takes five seconds to load images in and out of the viewer. Team driven efforts to alleviate such large time costs come with additional challenges due to inconsistencies in reporting and methods of performing QC

To address the above concerns (Fig. 1), we propose that an effective QC process for team-driven, large-scale neuroimaging datasets should meet the following design criteria: 1.) a consistent way to visualize the outputs of each image processing pipeline that is run on the data, 2.) a method for quick visualization and QC assessment in order to minimize the amount of time spent on the QC process without compromising the integrity of the QC process, and 3.) a manner by which QC results from different researchers can be easily and seamlessly aggregated across datasets and pipelines to provide a comprehensive QC of the entire database. Addressing these criteria in a QC pipeline can help ensure brevity and completeness in QC while providing a coherency of global QC across all datasets.

We propose a QC system for a database consisting of national-scale diffusion-weighted imaging (DWI) and T1-weighted (T1w) MRI that adheres to the aforementioned design criteria. Our database consists of 20 large-scale datasets with 16 distinct preprocessing and postprocessing pipelines. We visualize all outputs in the portable network graphics (PNG) format. For viewing the PNGs and assessing quality, we build an application using a Flask (https://github.com/pallets/flask) server that allows researchers to quickly look through PNGs and easily document when data processing has failed (Fig. 2). The documentation method also allows us to easily aggregate QC assessments across pipelines and datasets in a CSV format that can easily be shared with collaborators or other team members. Further, we provide insight into what expected outputs are for frequently used preprocessing and postprocessing algorithms we employ, as well as common instances when the algorithms fail to produce expected outputs, showing how our model can be easily adapted to other custom pipelines.

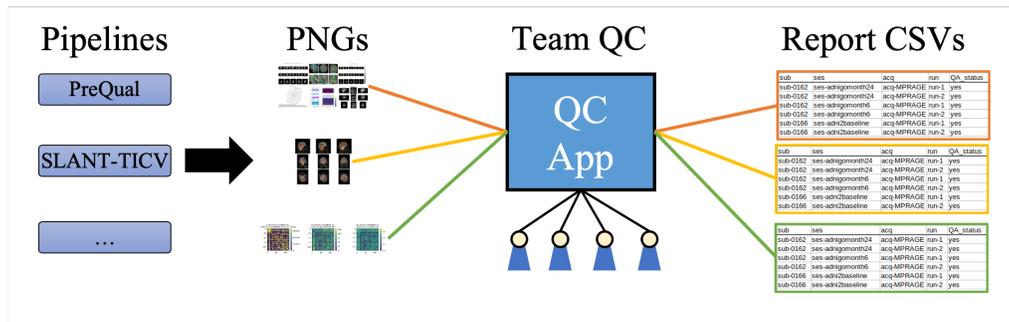

**Fig. 2** We propose an efficient and scalable method for standardized deep QC of neuroimaging pipeline outputs that works within a team setting. For each pipeline output, we provide a visualization as a PNG file. The PNGs are then loaded into a QC app that permits quick scanning of outputs for abnormalities and failures. Any team member can start an instance of the app, and results are standardized in a structured CSV file for each pipeline

## 2. METHODS

Upon receiving MRI data, we convert from the Digital Imaging and Communications in Medicine (DICOM) (Mildenberger et al., 2002) to Neuroimaging Informatics Technology Initiative (NIFTI) (https://nifti.nimh.nih.gov/nifti-1) format when data are provided as DICOM files; no conversion is required if data are only provided in the NIFTI format. For an initial rapid filtering of data, we initially ignore any DWI that do not have corresponding BVAL and BVEC files that indicate the direction and strength of the diffusion-weighting, asking data providers for missing files if possible. We ignore DWI with fewer than 6 volumes unless it is a reverse-phase encoding scan for susceptibility distortion correction that accompanies a more highly sampled DWI. We also visualize raw T1w and DWI in batches using *nibabel* (Brett, 2022) to load images and *matplotlib* (Hunter, 2007) to display and output them as PNGs to ensure that scans are the correct modality. We organize raw and derived data according to the Brain Imaging Data Structure (BIDS) (v.1.9.0) format as mentioned in our previous work (Kim et al., 2024).

### 2.1 Data Processing Pipelines

For processing raw MRI scans, we run 16 different pipelines in total across T1w and DWI modalities (Table 1). The pipelines have a variety of purposes, including preprocessing, signal modeling, image registration, segmentation, tractography, connectomics, and surface reconstruction. There exists a dependency chain for the pipelines, with many also requiring both T1w and DWI modalities (Fig. 3). Some pipelines, such as PreQual (Cai et al., 2021), already create QC documents designed to assess the quality of specific outputs. However, most times these pre-existing QC documents are not in a form that is optimized for high throughput quality checks.

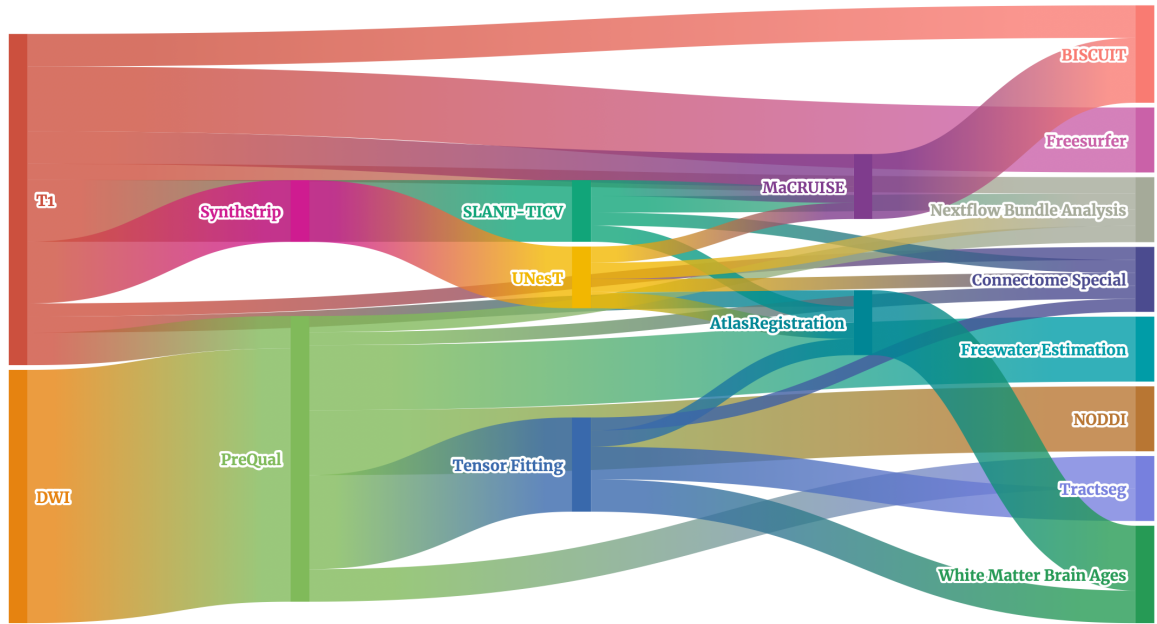

**Fig. 3** The 16 pipelines we run on our data form a complex dependency chain, and thus, it is essential to perform quality control on each step so that data are not misinterpreted during downstream analyses. Created with *flourish.studio*

**2.2 Generation of Quality Control Files**

The format we select for viewing outputs is PNG, since Joint Photographic Experts Group (JPEG) format is a lossy compression that does not retain all image information whereas portable document format (PDF) files can consist of multiple pages and have a large file size, making them slower to load. Some pipelines require generation of PNGs from outputs, whereas the atlas registration pipelines and the Connectome Special pipeline output a QC PNG that can directly be used in our process. PreQual (Cai et al., 2021), the Brain Shape Computing Toolbox (BISCUIT) (Lyu et al., 2017; Lyu, Kim, Girault, et al., 2018; Lyu, Kim, Woodward, et al., 2018), and Nextflow Bundle Analysis (https://github.com/MASILab/francois_special_spider.git) output multi-page QC PDFs as part of their respective pipelines, and so we combine these pages into a single PNG for more efficient QC. We automatically determine a process to be a failure for pipelines that do not produce all of their intended outputs, including the QC PDFs/PNGs when applicable.

For the remaining pipelines that require generation of a QC PNG, we use a variety of Python libraries to load and view images in a way that best allows us to perform efficient QC. We combine tensor fitting and atlas registration as a single QC step, overlaying the atlas label map on a fractional anisotropy (FA) map that is extracted from the DWI tensor fit using *nibabel* and *matplotlib* for a tri-planar view at three slices. We similarly overlay the segmentation outputs from UNesT (Yu et al., 2023, 2024), SLANT-TICV (Y. Liu et al., 2022), and MaCRUISE (Huo et al., 2016) over the T1w image used as input. The free water estimation pipeline QC step has side-by-side triplanar views of the uncorrected and free water corrected FA map in three slices. To visualize each of the individual bundles for Tractseg (Wasserthal et al., 2018), we use *scilpy* (https://github.com/scilus/scilpy) to overlay the bundles on tri-planar views of the FA map. For Synthstrip (Hoopes et al., 2022), we simply visualize the skull-stripped T1w image for all three planes. For NODDI (Zhang et al., 2012), we visualize all tissue parameter maps in three axial slices. Finally, the *fsqc toolbox* (https://github.com/Deep-MI/fsqc.git) was used to generate QC PNGs for cortical surface segmentations of the brain. All PNG files are organized separately from the raw and processed data in a QC archive.

**2.3 Performing Quality Control for a Dataset and Pipeline**

Once we have generated the PNG files for a pipeline that has been run on a dataset, we visually inspect each file using a QC app running on a Python backend with Flask (Fig. 4). Each researcher can start their own instance of the QC app and select the dataset and pipeline that they wish to perform QC on. PNG files are then pulled from the QC database and loaded in the QC tool for the user to view. Whenever a failed pipeline is detected, the user can document that the output is a failure, and the results will be pushed in real-time to a database of the QC results for that pipeline (Fig. 5). The source code for the app can be found here: (https://github.com/MASILab/ADSP_AutoQA.git).

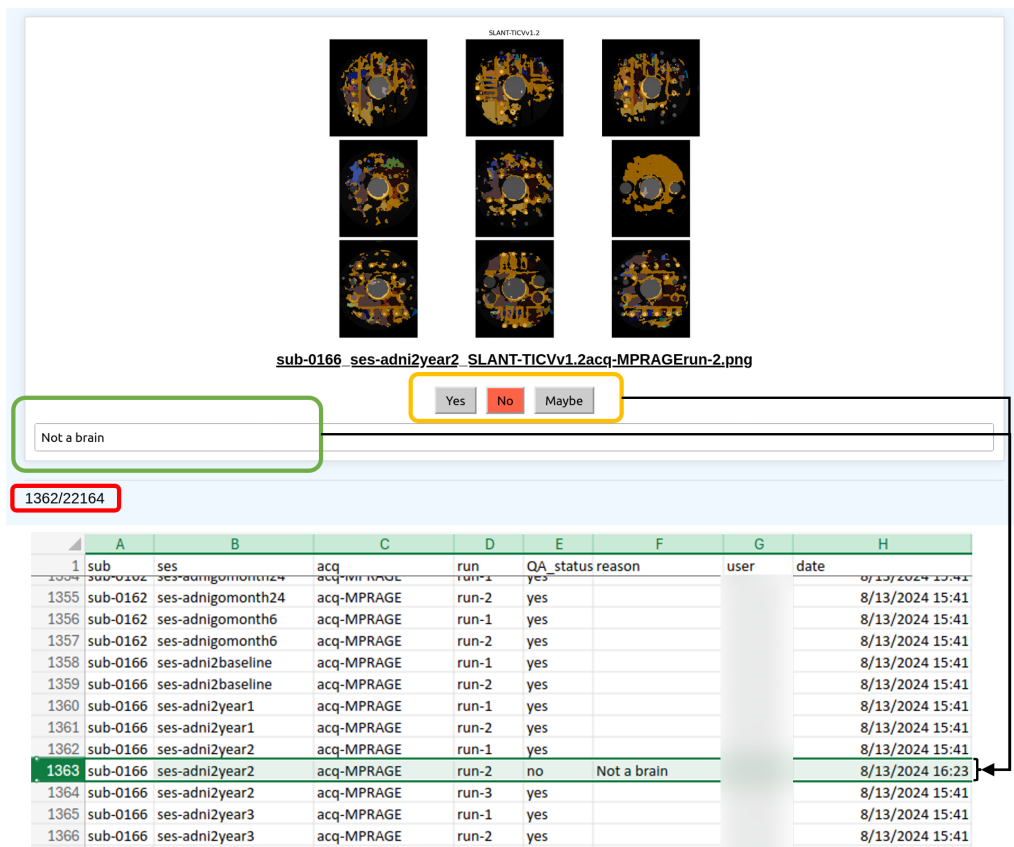

**Fig. 4** The proposed QC tool enforces consistent visualization of pipelines outputs through uniformly generated QC PNG documents for each respective pipeline. The homogenous format of the documents and high success rate for pipelines allow QC users to very quickly cycle through PNGs to catch any abnormalities. Moving through PNG documents can be done either manually with the arrow keys or automatically through the montage feature of the app (not shown). A counter is maintained (red box) so the user can know how many documents are left to view. Most outputs are expected to be good, and thus all QC results are initialized as "Yes" (passes QC) in order to minimize the manual effort in reporting. If the output is not satisfactory, the user can click the corresponding button to indicate their decision (yellow box) and articulate the reason for their verdict with some accompanying text (green box). Any changes are automatically pushed to a CSV document that maintains structured information about the results of the QC for a pipeline and dataset. All QA CSV results documents are formatted the same way and can thus be easily merged across pipelines and datasets so that QC decisions can be shared among team members and collaborators in an easily parseable format

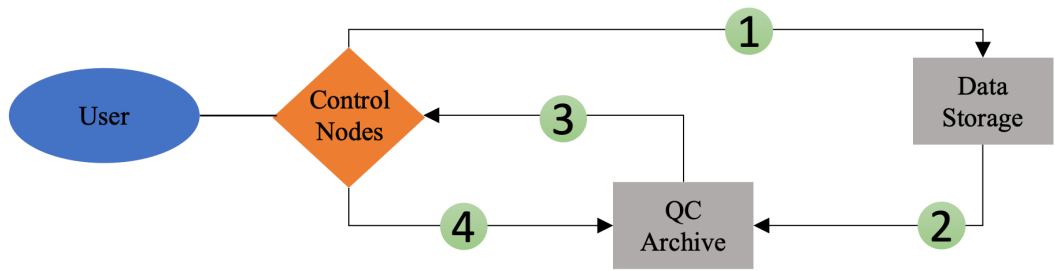

**Fig. 5** Our method allows for an efficient quality control pipeline: 1.) QC files are generated by querying the data storage from control nodes 2.) and then stored on an archive separate from the data. 3.) To perform QC, files are pulled from the QC archive 4.) and then QC results are pushed back in real time for other researchers to use

## 3. ADHERENCE TO DESIGN CRITERIA

### 3.1 Consistent Visualization

In order to ensure a single consistent visualization practice across all researchers, we perform all QC of pipeline outputs using the Python QC app, with all QC files having the same PNG image format. Thus, there is no variety in the manner in which outputs are viewed, preventing any preferences for tools or image viewers from causing variations in the QC process. Further, we use the same method for each pipeline to create QC PNGs across all datasets, enforcing all members of the QC team to visualize each respective pipeline output in the same way. This method also makes expectations of proper pipeline outputs homogenous across our QC team.

### 3.2 Efficient and Proper Quality Control Minimizing Time and Effort

As visually looking at every pipeline output is of utmost importance for a deep QC process, we ensure that all outputs are viewed by a QC team member. As mentioned in Section 2.2, the PNG format is a single image, allowing for fast loading times and easier visualization unlike PDFs, while still maintaining a high image quality, unlike lossy compression formats such as JPEG. The aspect of the QC process that best increases time efficiency without sacrificing quality of QC is the Python QC app, which permits users to quickly cycle through PNGs for an entire pipeline of a dataset. The app pre-loads the images into memory prior to visualization, permitting incredibly fast refresh speeds on the user's monitor. To further reduce the effort of the QC process, the app allows the user to automatically montage the PNGs at either a faster or slower speed depending on the preference. If there is an output that the user finds issue with, they may stop the montage at any time to record a failed QC instance. The user also has freedom to cycle back and forth between images in case they would like to return to a previously viewed output.

Another point of efficiency is the minimal effort of reporting QC results. As we expect most outputs to be properly processed data, the status of all outputs is initially set to have passed QC, minimizing the amount of manual effort in reporting. If there is a bad output encountered, reporting the status requires only clicking the button that indicates a failed output. If further explanation is needed, there is also the option of entering a small text blurb to accompany the QC decision. Finally, after the user has reported the QC status, any detected change is automatically pushed to the archive for other users to see.

### 3.3 Consistent Reporting and Aggregation of Results

To ensure a uniform QC reporting format, the QC app permits users to report status as one of three options. "Yes," means that the pipeline has properly run on the input data. "No," means that there is a failed instance of the algorithm in some manner, where the output is an unexpected result. Lastly, "Maybe," is a more nuanced decision that indicates that the algorithm performance is within expectation or believable, but the outputs are not visually good results. For example, a "Maybe" instance could be if a tractography algorithm produced a very small number of streamlines for a white matter bundle that are anatomically correct, but the bundle is not as thick as it could be. A "Maybe" is intended to leave the decision up to the researcher who is using the data for their analysis. Should there be any additional need for clarification, the user may also add some text to explain their QC decision. Thus, there is no ambiguity and little subjectivity in the reporting of results, as opposed to methods such as a rating scale.

For each dataset and pipeline, the QC results are also stored in a consistent format as a CSV file, with a row for each individual output for a pipeline run on the dataset. The columns indicate identifiers, such as the BIDS tags for the corresponding data, as well as all the QC information, including the QC status, user, time of the QC report, and any notes left by the QC reporter. As all QC reports are stored in this consistent CSV format, the results are simply aggregated together with a single script to create a QC document that encompasses the results across all pipelines of a dataset. The CSV is easily readable and can be shared with any other team members or collaborators who use the data.

Table 1. A list of all the processing pipelines that are run on our datasets.

| **Pipeline** | **Modality** | **Description** |
| --- | --- | --- |
| PreQual (Cai et al., 2021) | DWI | A preprocessing pipeline for DWI data that includes denoising and correction for motion, susceptibility-induced, and eddy-current induced artifacts as well as slice-wise signal dropout imputation. If multiple images are provided, inter-scan normalization is performed as well. May require a T1w image if no reverse phase encoding scans are available. Also outputs a PDF for quality checks. |
| Tensor Fitting (Tournier et al., 2019) | DWI | Uses MRtrix3 algorithms to model the diffusion signal as a tensor from preprocessed DWI data (*dwi2tensor*) with volumes <= 1500 and then subsequently extract tensor feature maps for FA, MD, AD, and RD (*tensor2metric*). |
| EVE3 Atlas Registration (Oishi et al., 2009, 2010; Tustison et al., 2014) | DWI/T1w | Using ANTs SyN registration, aligns the JHU template to a preprocessed DWI scan. Also calculates the calculate the mean, median, and standard deviation of the tensor scalar maps for each region in the EVE3 atlas. Requires a T1w image from the same scanning session and uses SLANT-TICV or UNesT segmentation for a brain mask. (https://github.com/MASILab/AtlasToDiffusionReg.git) |
| MNI152 Atlas Registration (Fonov et al., 2009; Tustison et al., 2014) | DWI/T1w | Affine and deformable registration of the preprocessed DWI to the MNI 152 template using ANTs SyN. Requires a T1w image from the same scanning session and uses SLANT-TICV or UNesT segmentation for a brain mask. (https://github.com/MASILab/AtlasToDiffusionReg.git) |
| NODDI (Zhang et al., 2012) | DWI | Uses a scilpy script to fit the NODDI model to preprocessed DWI data and export scalar maps of the tissue parameters for the model. Only applicable to multi-shell DWI data. (https://github.com/scilus/scilpy) |
| Nextflow Bundle Analysis (Côté et al., 2015; Cousineau et al., 2017; DI Tommaso et al., 2017; Garyfallidis et al., 2018; Rheault, 2020, 2021; Rheault et al., 2021; Theaud et al., 2020; Yeatman et al., 2012) | DWI/T1w | A fusion of four separate pipelines using Nextflow: Tractoflow, RecobundlesX, Tractometry Flow and Connectoflow. Nextflow generates whole-brain tractography, RecobundlesX is a multi-atlas segmentation of white matter bundles, Tractometry Flow segments whole-brain tractography into 36 bundles and calculates average tensor scalar values within entire bundles and subsections, and Connectoflow generates connectomics matrices. Requires a T1w image from the same scanning session and uses SLANT-TICV or UNesT segmentation. |
| Connectome Special (Rubinov & Sporns, 2010; Tournier et al., 2019) | DWI/T1w | Constructs a structural connectome of the brain and derived complex graph measures from preprocessed DWI using a combination of MRtrix3 tools and *scilpy* scripts. Requires a T1w image from the same scanning session and uses SLANT-TICV or UNesT segmentation. |
| Tractseg (Wasserthal et al., 2018) | DWI | Segments 72 unique white matter bundles using the Tractseg pipeline. Taking the Tractseg bundles, then uses *scilpy* scripts to calculate the mean and standard deviation of tensor scalars within each bundle-defined region of interest. |
| Free water Estimation (Pasternak et al., 2009) | DWI | Uses the Pasternak et al. method of free water estimation and extraction from a diffusion tensor model. |
| White Matter Brain Ages (Gao et al., 2024) | DWI | Gives estimates of the white matter age from FA and MD scalar maps in the MNI space both with and without minimizing the use of macrostructural information. |
| SLANT-TICV (Landman (ed.) & Warfield (ed.), 2019; Y. Liu et al., 2022) | T1w | A deep learning-based segmentation of the brain into 133 labels under the BrainCOLOR protocol in addition to two labels for the intracranial vault and posterior fossa. |
| SynthStrip (Hoopes et al., 2022) | T1w | Creates a brain mask from a T1w image. Only used when T1w images are defaced, as defacing methods impact results of segmentation algorithms that expect a non-altered T1w image. |
| UNesT (Yu et al., 2023, 2024) | T1w | Used for BrainCOLOR parcellation of a skull-stripped T1w image, as a separate model was trained for segmentation of skull-stripped images. |
| MaCRUISE (Huo et al., 2016) | T1w | Refinement of the BrainCOLOR labels from SLANT-TICV or UNesT based on a cortical surface reconstruction of the brain. |
| BISCUIT (Y. Liu et al., 2022; Lyu et al., 2017; Lyu, Kim, Girault, et al., 2018) | T1w | Performs cortical surface reconstruction and estimates curvature, folding, and shape measurements. |
| Freesurfer (Fischl, 2012) | T1w | Uses the recon-all tool to perform cortical reconstruction of a T1w image with the Freesurfer toolkit. |

## 4. RESULTS

To compare the efficiency of our deep QC method to deep QC using the FSLeyes image viewer, we perform QC of 200 SLANT-TICV segmentation outputs from the Alzheimer's Disease Neuroimaging Initiative (ADNI) (Weber et al., 2021) with both methods. We find using a script to open output segmentation in FSLeyes overlayed on a corresponding T1w image takes roughly 20 minutes to QC 200 outputs when there are no failed instances, where most time is spent loading in the images. Using our QC app, the same task takes only about 50 seconds using the montage function when there are no failed instances, highlighting the efficiency of our method. Furthermore, the automated tracking of failed outputs and easy aggregation of results reduces the amount of time and effort to manually record failed instances and combine QC efforts across our team.

### 4.1 PreQual Quality Control

When performing QC on the PreQual preprocessing pipeline for DWI data, there are a several outputs that we focus on to determine if there is an issue with the data, since PreQual itself is an amalgamation of several different algorithms and tools.(Cai et al., 2021) As PreQual focuses on ensuring the data are properly formatted, the QC mainly revolves around the original data quality and associated metadata needed to run the pipeline.

First, we look at the page detailing the outputs of the EDDY tool from FSL and intensity across volumes. For any 3D volume in a 4D DWI image, the intensity is related to the biological environment of the tissue and the gradient applied for diffusion-weighting (Derek K Jones, 2010). As the signal intensity $S$ at any voxel in the image is modeled as:

$$S = S_0 e^{-bg^T D g} \qquad (1)$$

where $S_0$ is the signal intensity without diffusion-weighting, $b$ is referred to as the b-value that represents the strength of the signal attenuation, $g$ is the direction of the applied gradient, and $D$ is the coefficient for the self-diffusion of water. Thus, the median intensity within a 3D volume should decay exponentially with respect to the b-value. We also expect there to be some amount of motion during the scan, and thus, plots of estimated translation and rotation of the head over time should be relatively smooth between adjacent volumes. For the EDDY slice-wise imputation algorithm, slices detected as outliers are replaced with a "prediction" by the algorithm (Andersson et al., 2016). Finally, we examine the chi-square analysis of the observed signal compared to the signal upon reconstruction from a modeled diffusion tensor, with the expectation that volumes with a b-value equal to or less than 1500 should be well-approximated by the tensor model (Jones et al., 2013).

Next, we look at the tensor visualization and sampled b-vector orientations, where b-vector is a term that refers to the unit normalized direction vector of the applied gradient pulse for diffusion weighting. While modeled diffusion tensors are visualized for every voxel, any errors can be easily seen when focusing on two areas: the corpus callosum for an axial slice and the corticospinal tract for a coronal slice. Axially viewed, the corpus callosum tensors should create a "U" shape and be oriented left-to-right (Fig. 6), whereas the corticospinal tract tensors should mostly run superior-to-inferior on a coronal slice view. Additionally, PreQual will perform a permutation check of the b-vectors to determine if there is an order and sign permutation that provides a more optimal streamline propagation for tractography along the direction of the diffusion tensor orientation. If the b-vectors are properly oriented, then the optimal orientation will be the original.

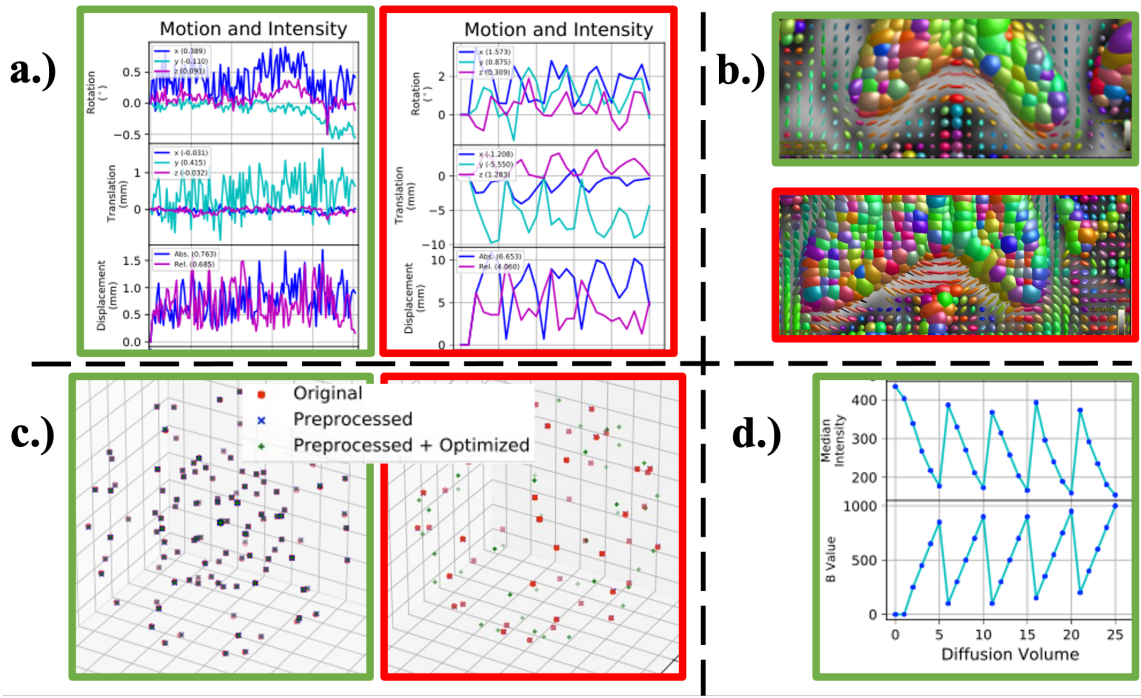

**Fig. 6** (a., green) While motion in DWI scans is not uncommon, we expect fluctuations between adjacent volumes to be relatively small. (a., red) However, an indication that the EDDY algorithm has failed is when intervolume motion is consistently on the magnitude of several millimeters . Properly oriented b-vector files will yield tensors whose directions point along white matter tracts, such as (b., green) the corpus callosum, and (c., green) will already have the optimal orientation. Any orientation issues will result in (b., red) the tensors pointing in the wrong direction or (c., red) will be determined as suboptimal when permuting the axes. (d.) The expectation for median volume intensity is roughly exponentially decreasing with increasing b-values

When there is no reverse phase-encoding scan available to perform susceptibility-induced distortion correction with TOPUP (Andersson et al., 2003), PreQual uses a deep-learning method called Synb0-DisCo (Schilling et al., 2019) to generate a synthetic 3D image from a paired T1w scan that can be supplied to TOPUP to correct the distortions. For these instances, we also visualize the synthesized image to ensure that it looks anatomically accurate for use in correction. Finally, we also visualize an FA map created from a tensor using all image volumes, where we expect the white matter to appear bright compared to the rest of the brain.

If there is any issue with the metadata of a scan, for instance, improperly reported b-values or b-vectors, the discrepancies are often times common for the entire dataset. Thus, for data preprocessing, we advocate a few preliminary pipeline runs with QC prior to parallelization of preprocessing for the entire dataset. The most frequently encountered issue with this regard is a flip of the reported b-vectors across the x-axis, which can be easily verified from the tensor orientation of the corpus callosum and comparison to the optimal b-vector orientation (Fig. 6). Another error to look out for is an improper synthetic image from Synb-DisCo, where the synthesized image has anatomical inaccuracies or is not a full brain (Supplementary Figs. S1, S2). For either of these examples, the subsequent issues are also present in the QC analysis. An indicator of a failed EDDY process when the estimated motion is hallucinated by the algorithm, resulting in translation or rotation plots where motion is sporadic and incredibly discontinuous (Fig. 6). Less frequent examples of errors are improper b-value reporting, which can be easily seen by comparing median intensity and b-value across volumes, and high chi-square values across multiple center slices of the brain for all volumes under 1500 b-values. While the FA map and slice imputation analysis help to provide additional assurance of a failed instance, issues with these two QC steps most often present with other issues rather than alone.

### 4.2 Segmentation Quality Control

For SLANT-TICV, UNest, or MaCRUISE, which are all segmentations focusing mostly on the gray matter of the brain, an overlay with slight opacity for the label map is usually indicative of success. The expectation is that the cortex of the brain is properly parcellated, with all labels in their approximate corresponding positions. (Fig. 7) While failures are very infrequent, the most common errors that do occur are when gray matter labels leak outside of the brain or into the white matter and vice versa. While SynthStrip is not a segmentation algorithm, its only place in the pipeline is to provide input to UNesT for defaced T1w images, so we include the QC of SynthStrip as part of the UNesT QC step.

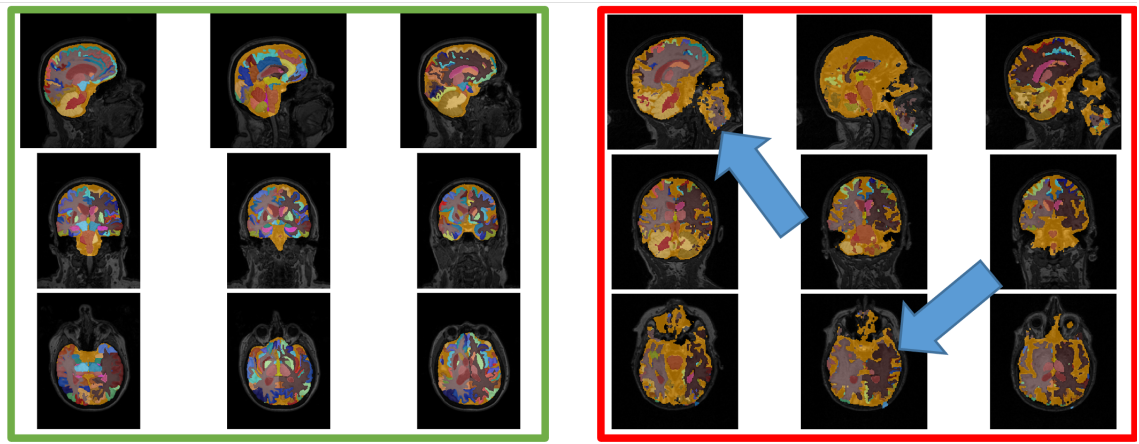

**Fig. 7** (Left) A good SLANT-TICV segmentation result has the TICV label (orange in figure) encapsulating the whole intracranial vault, with the cortical substructures appropriately labeled as well. (Right) Most common improper segmentations have labels outside of the brain or incorrectly segmented regions. MaCRUISE and UNesT QC documents and processes are similar to those of SLANT-TICV

### 4.3 Tensor Fitting and Atlas Registration

For registration of the JHU template, we visualize the transformed EVE3 labels in the subject space on top of an FA map extracted from a tensor fit using volumes with b-values less than 1500. For the MNI 152 registration, we visualize the MNI 152 template and FA/MD maps, but with affine and deformable transformations applied to place them in MNI 152 space. Both have been masked with the binarized SLANT-TICV/UNesT T1w image segmentation. We expect the registration to result in the EVE3 labels to roughly align with the white matter tracts present on the FA map (Fig. 8). As for the MNI registration, we expect the macrostructural information to be minimized for the deformable transformation and relative alignment for the affine only registration (Fig. 9). Failures are usually very apparent, with the labels or applied mask being wildly off-center from the FA map.

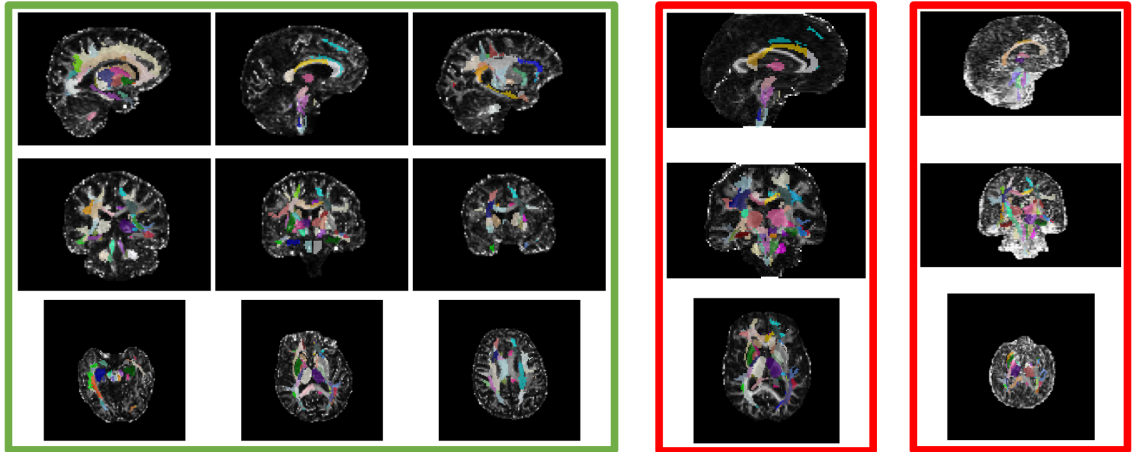

**Fig. 8** (Left) Proper registration of the EVE template into the DWI data space results in an alignment of segmented white matter pathways from the annotated EVE3 label map. Additionally, a tensor fit is considered to pass QC if the white matter tracts show as bright compared to the rest of the brain. (Middle) A poor registration can be seen when the labels do not align with the white matter tracts, (Right) and a poor tensor fit can result in a very noisy FA map. All images are viewed in the native participant DWI space

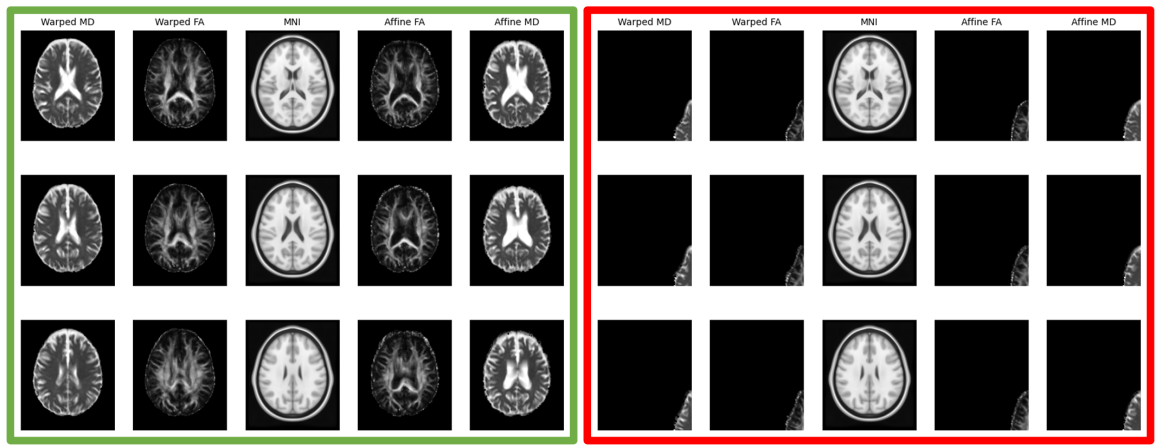

**Fig. 9** (Left) Proper MNI registration of DWI data results in a minimization of macrostructural features for the deformable registration and overall anatomical alignment for the affine registration. (Right) Severe failures for registration result in the brain can result in the brain being outside of the field of view. All images are viewed in the MNI space

### 4.4 Tractseg

As Tractseg only outputs 72 different white matter bundles, we visualize each bundle in a separate PNG. All tracts are expected to have a reasonable number of streamlines to give a "full" appearance to the white matter bundle (Fig. 10). Most encountered issues with outputs are a small number of streamlines that make the bundle appear wispy, with some bundles even having no streamlines to visualize. A complete lack of streamlines will be caught in later processing when calculating the average tensor scalar metrics within the bundle, as the calculation of the average will be infeasible, but the QC is important to identify such instances with only a handful. Much rarer issues are when streamlines are improperly located or terminate early.

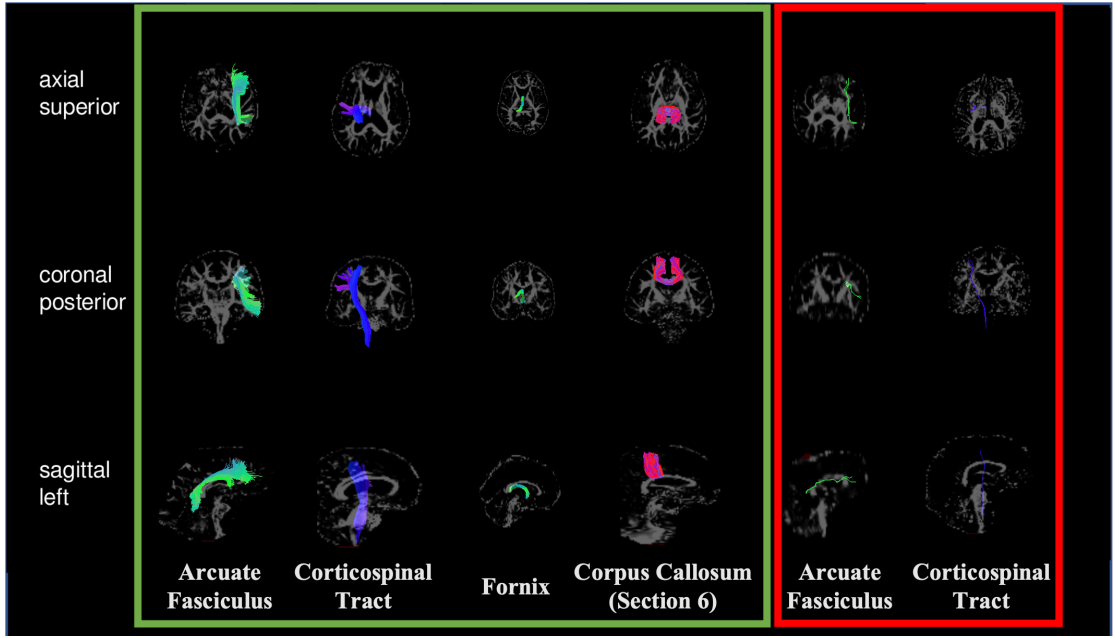

**Fig. 10** When Tractseg bundles are properly generated, there are a sufficient number of streamlines to give the bundle a full appearance (examples of major bundles in green). The most common failed outputs for Tractseg occur when there are only a handful of streamlines, giving the bundle a wispy appearance upon visualization (examples in red)

**4.5 Brain Shape Computing Toolbox**

Brain Shape Computing Toolbox (BISCUIT) QC consists of visualizing the cortical surface reconstruction parcellation as well as various cortical surface scalars mapped onto the brain surface. We expect the parcellation of the cortex to be mostly symmetric when comparing hemispheres, with segmented regions in the proper positions (Supplementary Fig. S3). The most indicative quantitative scalar map for QC is cortical thickness, which should be in a range of around 2-4 millimeters on average. For failed instances, the parcellation is either visibly asymmetric, the regions are not in their proper locations, or some regions are missing. Most failed outputs have cortical thickness maps that are much lower than the expected range.

**4.6 Freesurfer**

Freesurfer QC visualization is an overlay of the segmentation maps and gray matter surface contour lines on the T1w image that was used as input. We expect the segmentations to align with their corresponding structures, for instance, the ventricles, the white matter map, the cerebellum, etc. (Supplementary Fig S4) The boundary between the gray and white matter segmentations should also be near the cortex. In most instances when the Freesurfer algorithm is unsuccessful, it fails to generate all the outputs, so the QC PNG cannot be made.

**4.7 Connectome Special**

The main output of the connectome special, i.e. the structural connectome matrix, is the focus of the QC for the pipeline outputs. We expect the connectome to be diagonally symmetric, with a reasonable number of strong connections between brain regions. There should be a very strong connection along the main diagonal for the connectome weighted by the number of streamlines (NOS). Additionally, the connectome weighted by the average FA value should have a relatively homogenous intensity (Fig. 11). Most issues that occur present when the main diagonal is not visible for the NOS-weighted connectome or if the FA-weighted connectome has large regions of inhomogeneity.

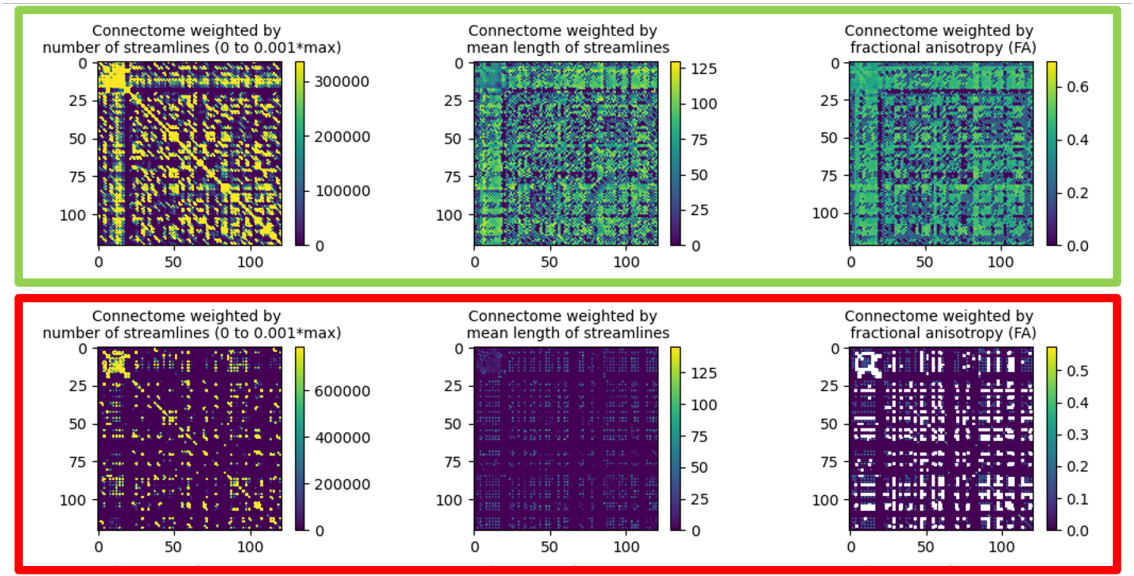

**Fig. 11** (Top) For properly generated structural connectome matrices, we expect there to be a strong main diagonal when weighted by the number of streamlines and a relatively homogenous intensity for connectomes weighted by mean FA. (Bottom) For failed outputs, the connectomes show sparse connectivity between brain regions, with many invalid connections for mean FA weighting and a weak main diagonal when weighting by number of streamlines

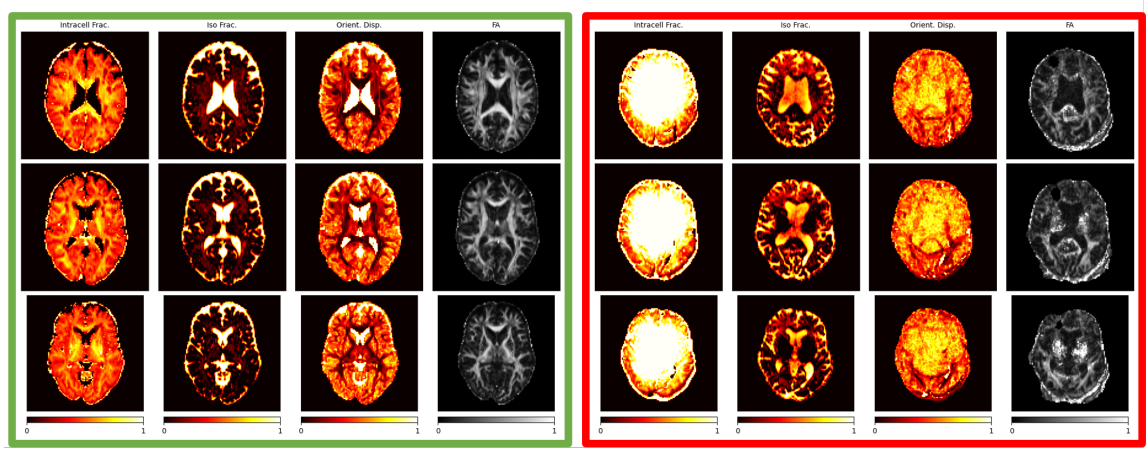

**Fig. 12** (Left) For NODDI outputs, we expect large white matter pathways (visible as high intensity for the FA map) to have a higher intracellular volume fraction, a low isotropic volume fraction, and a low orientation dispersion. In contrast, the CSF should have an intracellular volume fraction of zero and both isotropic volume fraction and orientation dispersion close to one. (Right) We consider NODDI outputs as failures if the estimated parameters do not adhere to these expectations

### 4.8 NODDI

For NODDI, we visualize axial slices for the estimated NODDI scalar measures of orientation dispersion index (ODI), isotropic volume fraction (ISOVF), intracellular volume fraction (ICVF), as well as an FA map for reference of white matter tracts. A high ODI value means that the orientations of axons are more disperse. Thus, very directed white matter tracts such as the corpus callosum should be dark. As ISOVF represents the fraction of cerebrospinal fluid for a voxel, we expect the CSF to be very bright. Finally, the ICVF is the fraction of voxels occupied by intracellular space, or axons and dendrites (collectively, neurites), as opposed to glial cells and cell bodies, so we expect the values in white matter to be larger than in the gray matter (Fig. 12). Most failures are obvious, with no anatomical structures visible on the NODDI maps.

### 4.9 Free water Estimation

For visualization of the free water estimation, we look at the free water-corrected FA maps compared to the uncorrected versions. We expect the free water correction to increase the relative intensity of the white matter pathways without altering their overall structure. A common failure is when the free water correction results in an FA map that is much noisier than the original or when the model fit results in an overestimation of the free water-corrected FA values in non-WM regions (Supplementary Fig. S5).

### 4.10 Nextflow Bundle Analysis

We visualize the whole brain tractography, the mean FA along the corpus callosum, arcuate fasciculus, and the corticospinal tract (also known as the pyramidal tract) with standard deviation value, and Connectomics matrices weighted by various quantitative metrics. The whole-brain tractogram is expected to have streamlines reaching every part of the cortex in the brain, with a large crossing fiber that connects the two brain hemispheres. We expect mean FA values along segmented bundles to range from 0.4 to 0.9, with values centered on average abound 0.5 (and slightly higher for the corpus callosum) (Supplementary Fig. S6). The Connectomics matrices, like with the Connectome Special, should have a reasonable number of connections. Like Freesurfer, in most instances when the algorithm is unsuccessful, it fails to generate a QC PDF.

### 4.11 White Matter Brain Ages

As the output for a white matter brain age is a single scalar value, there is no qualitative visualization of the outputs on an individual level.

## 5. DISCUSSION

For a deep QC of pipeline outputs, visual inspection of every output is paramount to ensure that no instances of failed algorithms are used in scientific analyses. As dataset size and the number of processes scale up, multiple researchers are needed to perform QC in a timely manner. We have demonstrated a proposed method of facilitating visual inspection that permits consistent and efficient QC across a team setting. The use of a unified QC app that enforces reporting of results in a specific manner promotes stability of the team QC. Additionally, the QC PNGs are generated in the same way for each individual pipeline. This predictability, coupled with the expectation of a proper pipeline output, increases efficiency by permitting rapid cycling through the QC PNG documents with the QC app. The standardized visualization of outputs also minimizes the amount of training required for new team members who have little to no experience with QC of the pipelines. Further, our automated documentation of QC results to the QC archive and ease in aggregating QC across pipelines reduces the manual effort for compiling the efforts made across the entire team. Combining results in an easily parsable CSV format also helps to expedite research and analysis for team members who use the data and for external collaborators. In these ways, we have successfully met the design criteria for a deep QC pipeline.

One of our main goals in using the PNG format to view outputs was to promote brevity and efficiency in the QC process. However, we do recognize that visualization with an image viewer that was built to render the datatype being visualized would be a more comprehensive level of QC, as a user would have more control in inspecting any part of high-dimensional data rather than a fixed viewpoint. As the QC design criteria were to assess outputs for failure of the algorithm, not a grading of quality level, we believe the decision to draw the line for the efficiency-thoroughness tradeoff should lie closer to the efficiency end. Thus, it was important to briefly visualize the data in a way that would maximize the amount of information in order to prevent caliber of the QC from being compromised. We understand the choice is subjective for what the best ways are to represent the data, and our decisions were based on preliminary QC efforts that we have built upon for the past two years. For the rare cases when there is still ambiguity in whether or not the algorithm failed, we do advocate visualization in a proper image viewer to better understand if the output is a successful result. We are not trying to discern why the algorithm may have failed in our pipeline, but if users wish to understand the algorithmic shortcomings, such visualization might help elucidate the issues.

We also note that our QC process is mostly focused on qualitative visualization of outputs rather than quantitative visualization. This is because quantitative visualization on an individual/session level would greatly reduce the efficiency of performing QC, as visualization of images is quicker to investigate than a series of numbers and is often more effective. For instance, a tractogram is a collection of lines in 3D space, where each line (or streamline) is composed of several $(x, y, z)$ coordinates. To quantitatively read the coordinates for millions of streamlines is an infeasible task. Representing the tractogram as a condensed quantitative metric would make the quantitative QC process easier, but then one would be relying on a derived measure instead of visualizing the data itself. As we continue performing QC on our data, we will consider including batch-level quantitative methods of QC on derived metrics, such non-parametric statistical methods like interquartile range (IQR) and median absolute deviation (MAD) for outlier detection (Mramba et al., 2024), in addition to our current pipeline in order to supplement the process. Such additions will be essential for pipelines like the white matter brain age that output only a single value.

Finally, in our approach, we are focused on identification of failed data processing. We do not attempt to correct data beyond any preprocessing that is common in the field for the imaging modality. We also are not trying to perform any iterative algorithmic fine-tuning to produce slightly higher quality outputs (for example, retraining a segmentation algorithm on more data and viewing the outputs for each new model). We suggest researchers who are interested in such algorithmic optimization to use a different method for QC of data, as our method is not intended for grading quality of outputs. We also do not attempt to assess how the quality of outputs from one pipeline influences the performance of a subsequent downstream analysis.

## CODE AND DATA AVAILABILITY - INFORMATION SHARING STATEMENT

The source code for the QC app can be found here: (https://github.com/MASILab/ADSP_AutoQA.git).

The datasets supporting the conclusions of this research are available, subject to certain restrictions. The datasets were used under agreement for this study and are therefore not publicly available. The authors may provide data upon receiving reasonable request and with permission.

Data used in the preparation of this article were in part obtained from the Alzheimer's Disease Neuroimaging Initiative (ADNI) database (adni.loni.usc.edu). The ADNI was launched in 2003 as a public-private partnership, led by Principal Investigator Michael W. Weiner, MD. The primary goal of ADNI has been to test whether serial magnetic resonance imaging (MRI), positron emission tomography (PET), other biological markers, and clinical and neuropsychological assessment can be combined to measure the progression of mild cognitive impairment (MCI) and early Alzheimer's disease (AD). (https://adni.loni.usc.edu/)

OASIS3 and OASIS4 data can be requested from https://sites.wustl.edu/oasisbrains/.

NACC data can be requested from https://www.naccdata.org/.

ROS/MAP/MARS are 3 harmonized prospective cohort studies of aging and dementia conducted by investigators at the Rush Alzheimer's Disease Center, Chicago allowing for efficient merging of data. All 3 studies were approved by an Institutional Review Board of Rush University Medical Center and all participants signed informed and repository consents. The 3 studies started in 1994, 1997, and 2004, respectively. Participants agree to annual clinical evaluation. Biennial brain imaging started in 2009. (Bennett et al., 2012; Fleischman et al., 2024; Marquez et al., 2020) Data can be requested at: https://www.radc.rush.edu/.

UK Biobank data can be requested from: https://www.ukbiobank.ac.uk/enable-your-research/apply-for-access.

HABS-HD data can be requested from: https://apps.unthsc.edu/itr/research.

WRAP data can be requested from: https://wrap.wisc.edu/data-requests-2/.

# ACKNOWLEDGEMENTS

This work was supported in part by the National Institute of Health through NIH awards K01-EB032898 (Schilling) and K01-AG073584 (Archer), grant number 1R01EB017230-01A1 (Landman), and ViSE/VICTR VR3029, UL1-TR000445, and UL1-TR002243. This work was supported by the Alzheimer's Disease Sequencing Project Phenotype Harmonization Consortium (ADSP-PHC) that is funded by NIA (U24 AG074855, U01 AG068057 and R01 AG059716). This work was conducted in part using the resources of the Advanced Computing Center for Research and Education (ACCRE) at Vanderbilt University, Nashville, TN. We appreciate the National Institute of HealthS10 Shared Instrumentation grant 1S10OD020154-01, and grant 1S10OD023680-01 (Vanderbilt's High-Performance Computer Cluster for Biomedical Research).

We thank Adam M. Saunders for his assistance in providing organized label maps for quality control measures.

Data collection and sharing for ADNI were supported by National Institutes of Health Grant U01-AG024904 and Department of Defense (award number W81XWH-12-2-0012). ADNI is also funded by the National Institute on Aging, the National Institute of Biomedical Imaging and Bioengineering, and through generous contributions from the following: AbbVie, Alzheimer's Association; Alzheimer's Drug Discovery Foundation; Araclon Biotech; BioClinica, Inc.; Biogen; Bristol-Myers Squibb Company; CereSpir, Inc.; Cogstate; Eisai Inc.; Elan Pharmaceuticals, Inc.; Eli Lilly and Company; EuroImmun; F. Hoffmann-La Roche Ltd and its affiliated company Genentech, Inc.; Fujirebio; GE Healthcare; IXICO Ltd.; Janssen Alzheimer Immunotherapy Research & Development, LLC.; Johnson & Johnson Pharmaceutical Research & Development LLC.; Lumosity; Lundbeck; Merck & Co., Inc.; Meso Scale Diagnostics, LLC.; NeuroRx Research; Neurotrack Technologies; Novartis Pharmaceuticals Corporation; Pfizer Inc.; Piramal Imaging; Servier; Takeda Pharmaceutical Company; and Transition Therapeutics. The Canadian Institutes of Health Research is providing funds to support ADNI clinical sites in Canada. Private sector contributions are facilitated by the Foundation for the National Institutes of Health (www.fnih.org). The grantee organization is the Northern California Institute for Research and Education, and the study is coordinated by the Alzheimer's Therapeutic Research Institute at the University of Southern California. ADNI data are disseminated by the Laboratory for Neuro Imaging at the University of Southern California. Data used in the preparation of this article were obtained from the Alzheimer's Disease Neuroimaging Initiative (ADNI) database (adni.loni.usc.edu). The ADNI was launched in 2003 as a public private partnership, led by Principal Investigator Michael W. Weiner, MD. The original goal of ADNI was to test whether serial magnetic resonance imaging (MRI), positron emission tomography (PET), other biological markers, and clinical and neuropsychological assessment can be combined to measure the progression of mild cognitive impairment (MCI) and early Alzheimer's disease (AD). The current goals include validating biomarkers for clinical trials, improving the generalizability of ADNI data by increasing diversity in the participant cohort, and to provide data concerning the diagnosis and progression of Alzheimer's disease to the scientific community. For up-to-date information, see adni.loni.usc.edu.

Research reported in this publication was supported by the National Institute on Aging of the National Institutes of Health under Award Numbers R01AG054073 and R01AG058533, R01AG070862, P41EB015922 and U19AG078109. The content is solely the responsibility of the authors and does not necessarily represent the official views of the National Institutes of Health.

The NACC database is funded by NIA/NIH Grant U24 AG072122. NACC data are contributed by the NIA-funded ADRCs: P30 AG062429 (PI James Brewer, MD, PhD), P30 AG066468 (PI Oscar Lopez, MD), P30 AG062421 (PI Bradley Hyman, MD, PhD), P30 AG066509 (PI Thomas Grabowski, MD), P30 AG066514 (PI Mary Sano, PhD), P30 AG066530 (PI Helena Chui, MD), P30 AG066507 (PI Marilyn Albert, PhD), P30 AG066444 (PI John Morris, MD), P30 AG066518 (PI Jeffrey Kaye, MD), P30 AG066512 (PI Thomas Wisniewski, MD), P30 AG066462 (PI Scott Small, MD), P30 AG072979 (PI David Wolk, MD), P30 AG072972 (PI Charles DeCarli, MD), P30 AG072976 (PI Andrew Saykin, PsyD), P30 AG072975 (PI David Bennett, MD), P30 AG072978 (PI Ann McKee, MD), P30 AG072977 (PI Robert Vassar, PhD), P30 AG066519 (PI Frank LaFerla, PhD), P30 AG062677 (PI Ronald Petersen, MD, PhD), P30

AG079280 (PI Eric Reiman, MD), P30 AG062422 (PI Gil Rabinovici, MD), P30 AG066511 (PI Allan Levey, MD, PhD), P30 AG072946 (PI Linda Van Eldik, PhD), P30 AG062715 (PI Sanjay Asthana, MD, FRCP), P30 AG072973 (PI Russell Swerdlow, MD), P30 AG066506 (PI Todd Golde, MD, PhD), P30 AG066508 (PI Stephen Strittmatter, MD, PhD), P30 AG066515 (PI Victor Henderson, MD, MS), P30 AG072947 (PI Suzanne Craft, PhD), P30 AG072931 (PI Henry Paulson, MD, PhD), P30 AG066546 (PI Sudha Seshadri, MD), P20 AG068024 (PI Erik Roberson, MD, PhD), P20 AG068053 (PI Justin Miller, PhD), P20 AG068077 (PI Gary Rosenberg, MD), P20 AG068082 (PI Angela Jefferson, PhD), P30 AG072958 (PI Heather Whitson, MD), P30 AG072959 (PI James Leverenz, MD).

This research has been conducted using the UK Biobank resource, application 16315.

Data were provided in part by OASIS - OASIS-3: Longitudinal Multimodal Neuroimaging: Principal Investigators: T. Benzinger, D. Marcus, J. Morris; NIH P30 AG066444, P50 AG00561, P30 NS09857781, P01 AG026276, P01 AG003991, R01 AG043434, UL1 TR000448, R01 EB009352. AV-45 doses were provided by Avid Radiopharmaceuticals, a wholly owned subsidiary of Eli Lilly.

Data were provided in part by OASIS - OASIS-4: Clinical Cohort: Principal Investigators: T. Benzinger, L. Koenig, P. LaMontagne.

Data contributed from MAP/ROS/MARS was supported by NIA R01AG017917, P30AG10161, P30AG072975, R01AG022018, R01AG056405, UH2NS100599, UH3NS100599, R01AG064233, R01AG15819 and R01AG067482, and the Illinois Department of Public Health (Alzheimer's Disease Research Fund). Data can be accessed at www.radc.rush.edu. More information about participant demographics and study information can be found here: https://www.rushu.rush.edu/research-rush-university/departmental-research/rush-alzheimers-disease-center/rush-alzheimers-disease-center-research/epidemiologic-research.

The data contributed from the Wisconsin Registry for Alzheimer's Prevention was supportedby NIA AG021155, AG0271761, AG037639, and AG054047.

We use generative AI to create code segments based on task descriptions, as well as debug, edit, and autocomplete code. Additionally, generative AI technologies have been employed to assist in structuring sentences and performing grammatical checks. It is imperative to highlight that the conceptualization, ideation, and all prompts provided to the AI originate entirely from the authors' creative and intellectual efforts. We take accountability for the review of all content generated by AI in this work.

## STATEMENTS AND DECLARATIONS

Any raw or derived data used in this paper were purely for visual representation in figures; there were no clinical analyses conducted on any data in this paper. Only 200 randomly selected scans from the ADNI dataset were used for the timing experiment described in the results section. We do not make any claims about how the efficacy of pipelines presented in this work function on data depending on the patient information or data acquisition methods.

The authors do not declare any competing interests related to this work.